\documentclass[3p,times]{elsarticle}

\usepackage{ecrc}

\volume{00}

\firstpage{1}

\journalname{Nuclear Physics A}

\runauth{Jan Fiete Grosse-Oetringhaus}

\jid{nupha}

\usepackage{amssymb}
\usepackage{lineno}

\usepackage[figuresright]{rotating}

\usepackage{units}
\newcommand{\bq}{\begin{equation}}
\newcommand{\eq}{\end{equation}}

\newcommand{\bqq}{\begin{eqnarray}}
\newcommand{\eqq}{\end{eqnarray}}

\newcommand{\cms}{\sqrt{s}}
\newcommand{\cmsNN}{\sqrt{s_{\rm NN}}}

\newcommand{\etain}[1]{$|\eta|$~$<$~$#1$}

\newcommand{\pt}{\ensuremath{p_{\rm{T}}}}

\newcommand{\pta}{\ensuremath{p_{\rm T, assoc}}}
\newcommand{\ptt}{\ensuremath{p_{\rm T, trig}}}

\newcommand{\Dphi}{\ensuremath{\Delta\varphi}}
\newcommand{\Deta}{\ensuremath{\Delta\eta}}
\newcommand{\Ntrig}{N_{\rm trig}}
\newcommand{\Nassoc}{N_{\rm assoc}}

\newcommand{\iaa}{\ensuremath{I_{\rm AA}}}
\newcommand{\raa}{\ensuremath{R_{\rm AA}}}

\newcommand{\figref}[1]{Figure~\ref{#1}}

\newcommand{\bfigFullPage}{\begin{figure} \begin{center} \vspace{0pt}}
\newcommand{\bfig}[1][t!]{\begin{figure*}[#1] \begin{center}}
\newcommand{\efig}{\end{center} \end{figure*}}
\newcommand{\btab}[1][t!]{\begin{table*}[#1] \begin{center}}
\newcommand{\etab}{\end{center} \end{table*}}

\newcommand{\eqref}[1]{(\ref{#1})}

\begin{document}

\begin{frontmatter}

%% Title, authors and addresses

%% use the tnoteref command within \title for footnotes;
%% use the tnotetext command for the associated footnote;
%% use the fnref command within \author or \address for footnotes;
%% use the fntext command for the associated footnote;
%% use the corref command within \author for corresponding author footnotes;
%% use the cortext command for the associated footnote;
%% use the ead command for the email address,
%% and the form \ead[url] for the home page:
%%
\title{Hadron Correlations Measured with ALICE}
%% \tnotetext[label1]{}
\author{Jan Fiete Grosse-Oetringhaus\fnref{fnlabel}}%\corref{cor1}}
\fntext[fnlabel]{for the ALICE collaboration}
\ead{jgrosseo@cern.ch}
%% \ead[url]{home page}
%% \cortext[cor1]{}
\address{CERN, 1211 Geneva 23, Switzerland}
%% \fntext[label3]{}

%\dochead{Hadron Correlations Measured with ALICE}
%% Use \dochead if there is an article header, e.g. \dochead{Short communication}

%% use optional labels to link authors explicitly to addresses:
%% \author[label1,label2]{<author name>}
%% \address[label1]{<address>}
%% \address[label2]{<address>}

\begin{abstract}
Angular particle correlations are a powerful tool to study collective effects and in-medium jet modification as well as their interplay in the hot and dense medium produced in central heavy-ion collisions. We present measurements of two-particle angular correlations of inclusive charged and identified particles performed with the ALICE detector. The near-side peak in the short-range correlation region is quantitatively analyzed: while the rms of the peak in $\phi$-direction is independent of centrality within uncertainties, we find a significant broadening in $\eta$-direction from peripheral to central collisions. The particle content of the near-side peak is studied finding that the $p/\pi$ ratio of particles associated to a trigger particle is much smaller than the one in the bulk of the particles and consistent with fragmentation of a parton in vacuum.
\end{abstract}

\begin{keyword}
heavy-ion collisions \sep two-particle correlations \sep near-side jet peak \sep jet broadening \sep proton over pion ratio 
%% keywords here, in the form: keyword \sep keyword

%% MSC codes here, in the form: \MSC code \sep code
%% or \MSC[2008] code \sep code (2000 is the default)

\end{keyword}

\end{frontmatter}

%%
%% Start line numbering here if you want
%%
\linenumbers

%% main text
\section{Introduction}
In central heavy-ion collisions at the LHC strong jet quenching has been reported by ALICE, ATLAS and CMS. The suppression of high-$\pt$ particles quantified by the nuclear modification factor $\raa$ drops as far as 0.14 \cite{aliceraa, cmsraa}. Further, a strong di-jet energy asymmetry has been reported \cite{atlasdijet, cmsdijet}, while the quenched energy reappears primarily at low to intermediate $\pt$ (\unit[0.5-8]{GeV/$c$}) and also outside the jet cone \cite{cmsdijet}. The measurement of the yield of particles associated to a high-$\pt$ trigger particle (\unit[8-15]{GeV/$c$}) quantified by $\iaa$ shows a suppression on the away side and a mild enhancement on the near side indicating that medium-induced jet modifications can also be expected on the near side \cite{aliceiaa}.

It is interesting to study the low to intermediate $\pt$ region where the quenched energy reappears with the aim of constraining jet quenching mechanisms. Further, measurements in this $\pt$ region allow to quantify interactions of high energetic partons and branched-off partons with the collectivity-dominated bulk.

Full jet reconstruction for jets with a transverse momentum of less than \unit[10-20]{GeV/$c$} is very difficult due to the large backgrounds \cite{jetbackground}. Two-particle angular correlations are a powerful alternative in this regime. This paper presents results from two-particle correlations which allow to extract a small signal over a large background stemming from collective effects and pure combinatorics.

\section{Detector \& Data Sample}
The ALICE detector is described in detail in \cite{alice}. The Inner
Tracking System (ITS) and the Time Projection Chamber (TPC) are used
for vertex finding and tracking.
The collision centrality is determined with the forward scintillators (VZERO) at $-1.7 < \eta < -3.7$ and $2.8 < \eta < 5.1$.
The main tracking detector is the TPC which allows reconstruction of good-quality tracks with a pseudorapidity coverage of \etain{1.0} uniform in azimuth. The reconstructed vertex
is used to select primary track candidates and to
constrain the $\pt$ of the track.
For particle identification (PID) the specific energy loss measured in the TPC as well as the time of flight measured by the TOF system is used.

In the presented analysis about 15 million minimum-bias Pb--Pb events recorded in fall 2010 at $\cmsNN = \unit[2.76]{TeV}$ as well as 55 million pp events from March 2011 ($\cms = \unit[2.76]{TeV}$) are used. 
These include only events where the TPC was fully efficient to ensure uniform azimuthal acceptance. Events are accepted which have a reconstructed
vertex less than $\unit[7]{cm}$ from the nominal interaction point in beam direction. Tracks are selected by
requiring at least 70 (out of up to 159) associated clusters in the TPC, and a $\chi^2$ per space point
of the momentum fit smaller than 4 (with 2 degrees of freedom per space point). In addition, tracks
are required to originate from within \unit[2.4]{cm} (\unit[3.2]{cm}) in transverse (longitudinal)
distance from the primary vertex.

The data is corrected for tracking efficiency and contamination by secondary particles using the HIJING \cite{hijing} and PYTHIA \cite{pythia6} event generators followed by particle transport and detector simulation based on GEANT3 \cite{geant3}.

\section{Two-Particle Angular Correlations}

\bfig
    \includegraphics[width=0.85\linewidth,clip=true,trim=0 12 0 11]{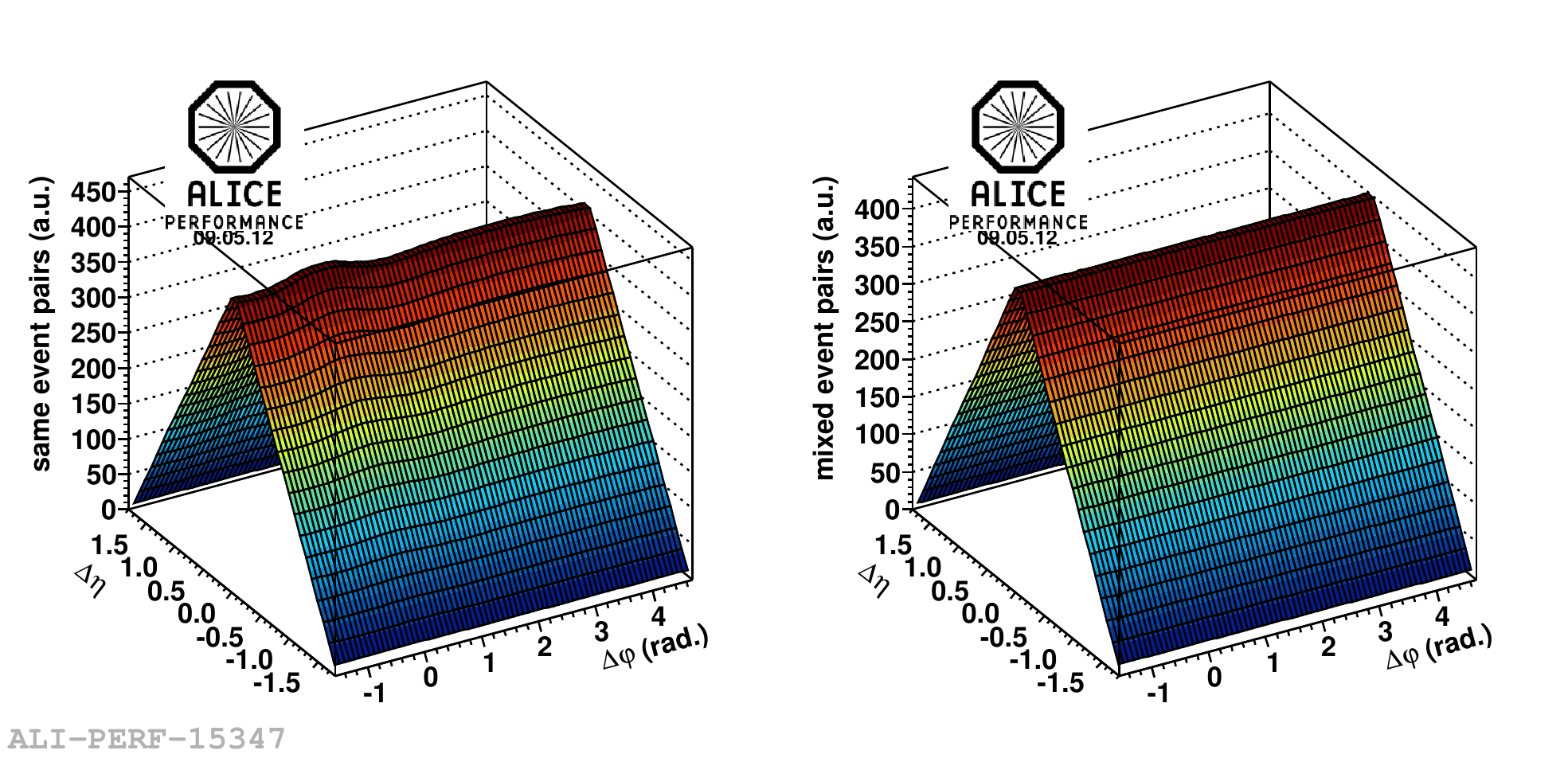}
    \caption{\label{samemixed} Correlation constructed from pairs of particles from the same event (left panel) and mixed events (right panel).}
\efig

The associated per trigger yield is measured as a function of the azimuthal angle difference $\Dphi = \phi_1 - \phi_2$ and pseudorapidity difference $\Deta = \eta_1 - \eta_2$:
\bq
  \frac{d^2N}{d\Dphi d\Deta}(\Dphi,\Deta) = \frac{1}{\Ntrig} \frac{d^2\Nassoc}{d\Dphi d\Deta} \label{etaphi_yield}
\eq
where $\Nassoc$ is the number of particles associated to a number of trigger particles $\Ntrig$. This quantity is measured for different ranges of trigger $\ptt$ and associated transverse momentum $\pta$ and in bins of centrality.

Two-track efficiency and acceptance are assessed by using a mixed-event technique: the differential yield defined in Eq.~\eqref{etaphi_yield} is also constructed for pairs of particles from different events. By definition all physical correlations are removed while those e.g. related to the acceptance remain. The events mixed with each other are selected with similar centralities and $z$-vertex positions. The angular correlation constructed from particles within the same event as well as with mixed events are presented in \figref{samemixed}. A typical triangular shape is obtained in $\Deta$ originating from the limited $\eta$-acceptance of the detector. Due to the uniform acceptance of the detector in $\phi$, no significant structures can be observed in \Dphi. Dividing those two distributions after proper normalization of the mixed-event distribution results in the per-trigger yield, see e.g. the left panel in \figref{subtraction}, the quantity which is studied in the following.

To correct for efficiency losses in case of particles being spatially very close in the detector volume a cut on the spatial separation in the active volume of the TPC is performed. Applying this cut consistently in the same and mixed event distributions recovers the particles lost due to the detector inefficiency.

\subsection{Near-Side Peak Shapes}

A typical per-trigger yield is shown in the left panel of \figref{subtraction}. Visible are the near-side peak concentrated around $\Dphi = \Deta = 0$ sitting on top of a ridge structure around $\Dphi = 0$ elongated in $\Deta$ whose origin is flow. On the away-side around $\Dphi = \pi$, a ridge mostly independent of $\Deta$ can be observed, consisting of the recoil jet peak and the modulation from flow. The per-trigger yield has quite a large (flat) pedestal compared to the signal modulations. 

We study the shape of the near-side jet peak by subtracting \Deta-independent effects. Those are estimated in the long-range correlation region at $|\Deta| > 1$ and subtracted from the region $|\Deta| < 1$. The center panel of \figref{subtraction} shows the projection to $\Dphi$ in $|\Deta| > 1$ (red) and $|\Deta| < 1$ (black). The difference between the two is the signal to be studied. By construction this procedure removes the away-side peak which is to a good approximation $\Deta$ independent. The right panel of \figref{subtraction} shows the subtracted per-trigger yield.

\bfig
  \includegraphics[width=0.33\linewidth,clip=true,trim=5 12 390 11]{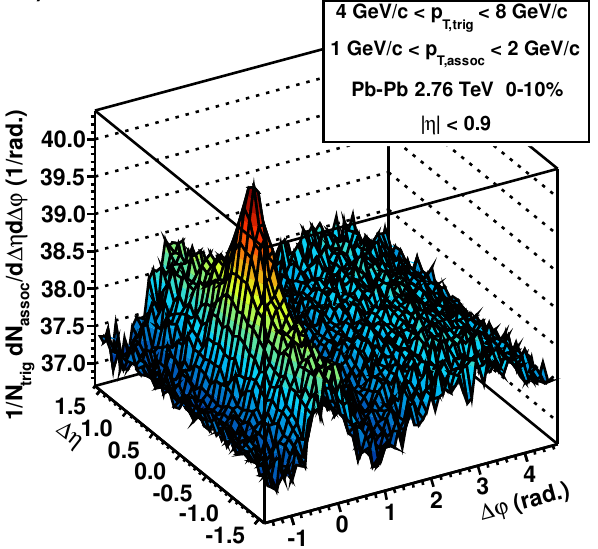}
  \hfill
  \includegraphics[width=0.33\linewidth,clip=true,trim=0 13 0 0]{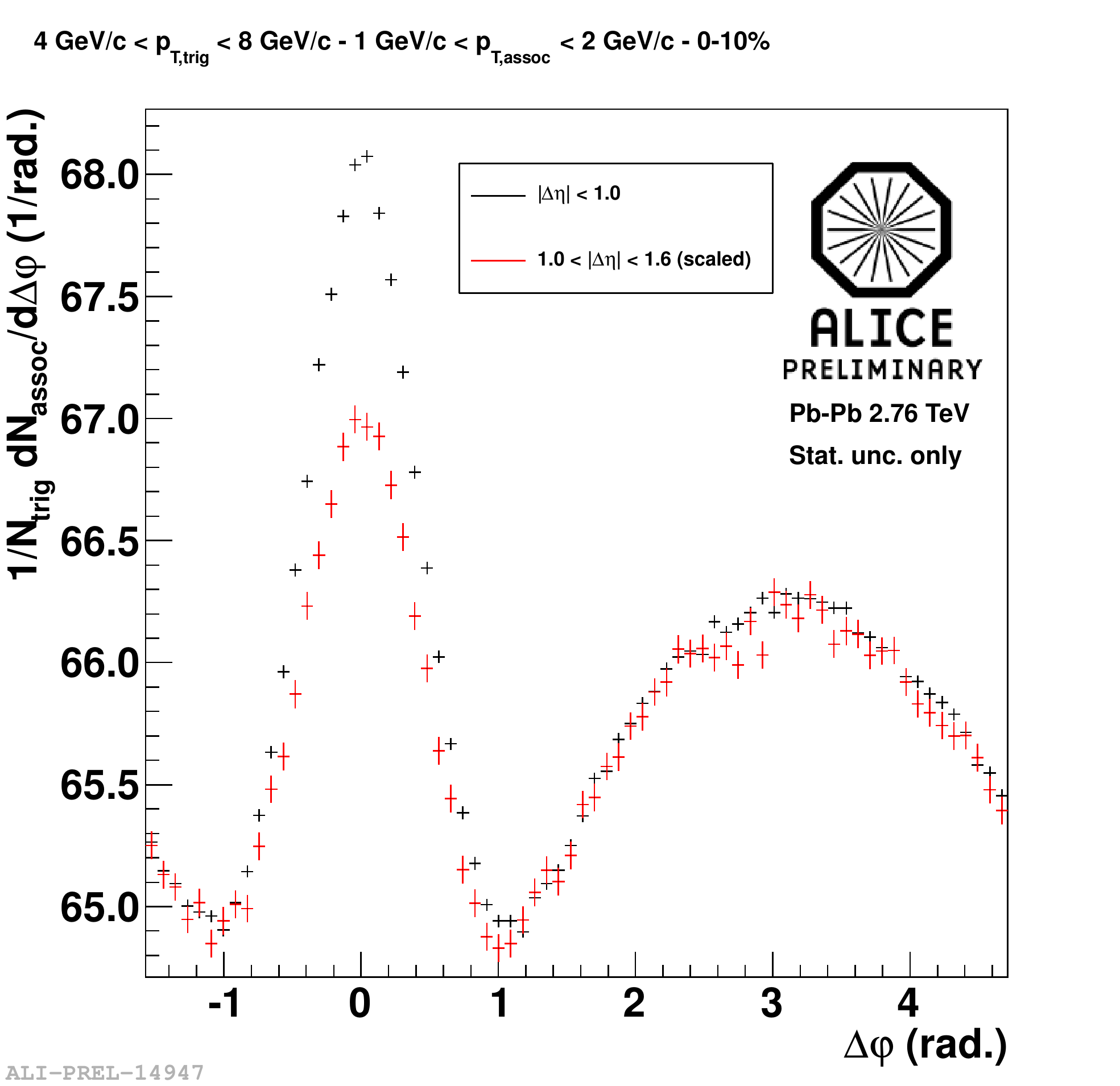}
  \hfill
  \includegraphics[width=0.33\linewidth,clip=true,trim=195 12 205 12]{2012-Jun-06-ex_2_1_0}
  \caption{\label{subtraction} Left panel: per-trigger yield; middle panel: projection to $\Dphi$ in $|\Deta| < 1$ (black) and $|\Deta| > 1$ (red); right panel: subtracted per-trigger yield.}
\efig

The near-side peaks are fitted with a superposition of 2 Gaussians which have their center at $\Dphi = \Deta = 0$. Such a fit function is chosen because it reproduces the features of the data in most bins ($\chi^2/ndf \approx 1.1 - 1.4$). We use the fit parameters to calculate the rms (equal to the variance, $\sigma$, for distributions centered at 0) of the distribution. In addition, we calculate the rms directly from the distribution as well as the excess kurtosis\footnote{Kurtosis $K = \mu_4 / \mu_2^2 - 3$; $\mu_n$ being the $n^{th}$ moment.} $K$ which is a measure of the peakedness of the distribution. 

\figref{rms} presents the rms in $\Dphi$ (left panel) and $\Deta$ (right panel) as a function of centrality; also shown are the results for pp collisions (shown at a centrality of 100). The rms in $\Dphi$ direction is rather independent of centrality while there is a significant increase in the rms in $\Deta$ direction towards central collisions. 
For peripheral and pp collisions, the rms is similar in $\Dphi$ and $\Deta$, e.g. about 0.4 for the lowest $\pt$ bin studied. It increases in $\Deta$ up to about 0.6 for 0-10\% centrality. A similar relative increase is seen for the other $\pt$ bins studied. Generally, the parameters continue smoothly from peripheral collisions to pp collisions. In \cite{jetflow} it was suggested that the interplay of longitudinal flow with a fragmenting high $\pt$ parton can lead to such an asymmetric peak shape. The lines in \figref{rms} are from models: for Pb--Pb AMPT (A MultiPhase Transport Code; version 2.25 with string melting) simulations \cite{ampt, ampttuning} which describe collective effects in heavy-ion collisions at the LHC reasonably well, are shown, while for pp PYTHIA 6.4 \cite{pythia6} with the tune Perugia-0 \cite{perugia0} is presented. These describe the rms in $\Dphi$ and $\Deta$ qualitatively and to some extent also quantitatively. AMPT models the interplay of a jet with flow with partonic and hadronic rescattering. \figref{kurtosis} shows the Kurtosis in $\Dphi$ and $\Deta$ direction. It increases from central to peripheral collisions: the near-side is less peaked towards central collisions. Also the Kurtosis is well described by AMPT and PYTHIA.

\bfig
  \includegraphics[width=0.48\linewidth]{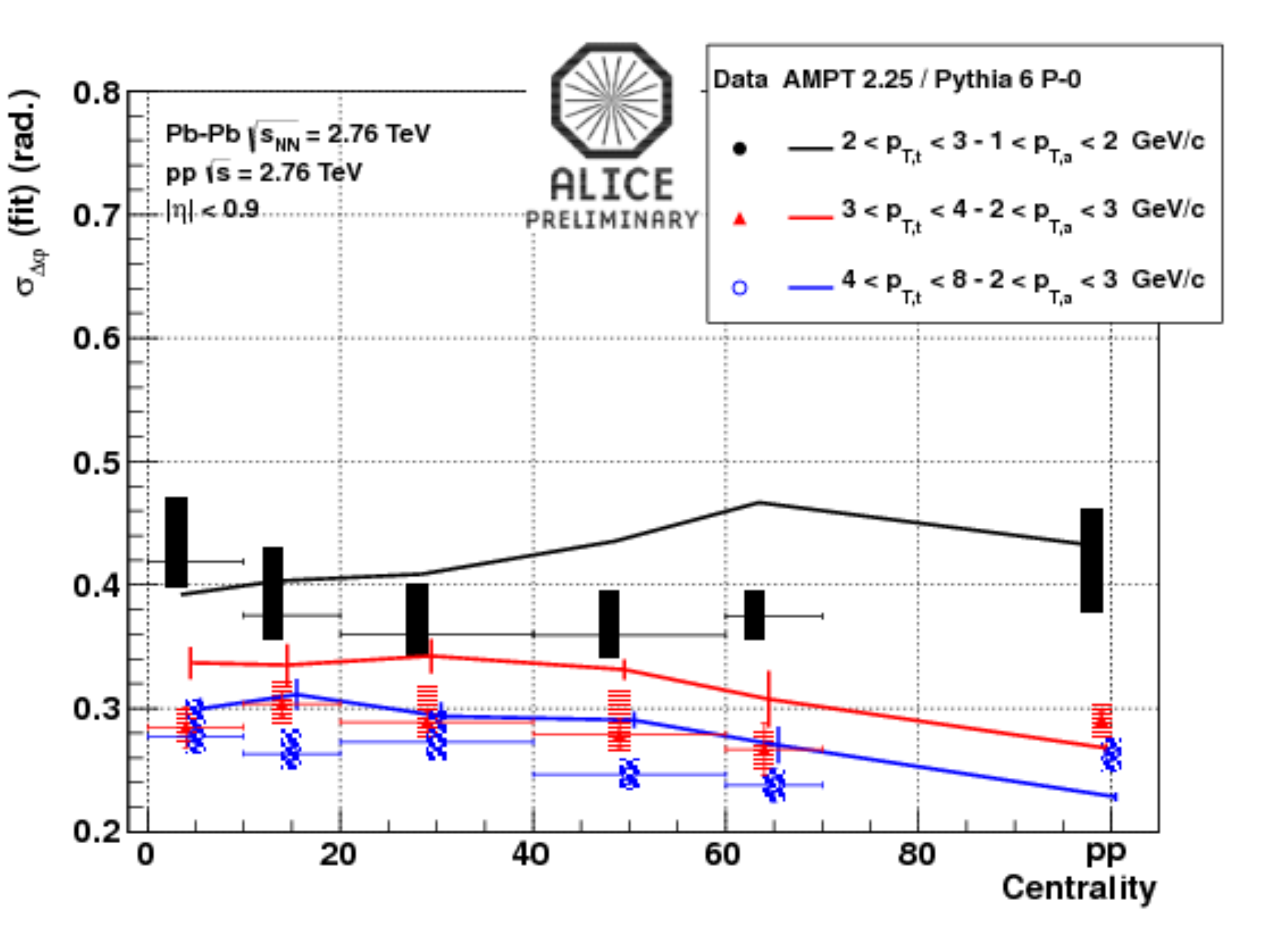}
  \hfill
  \includegraphics[width=0.48\linewidth]{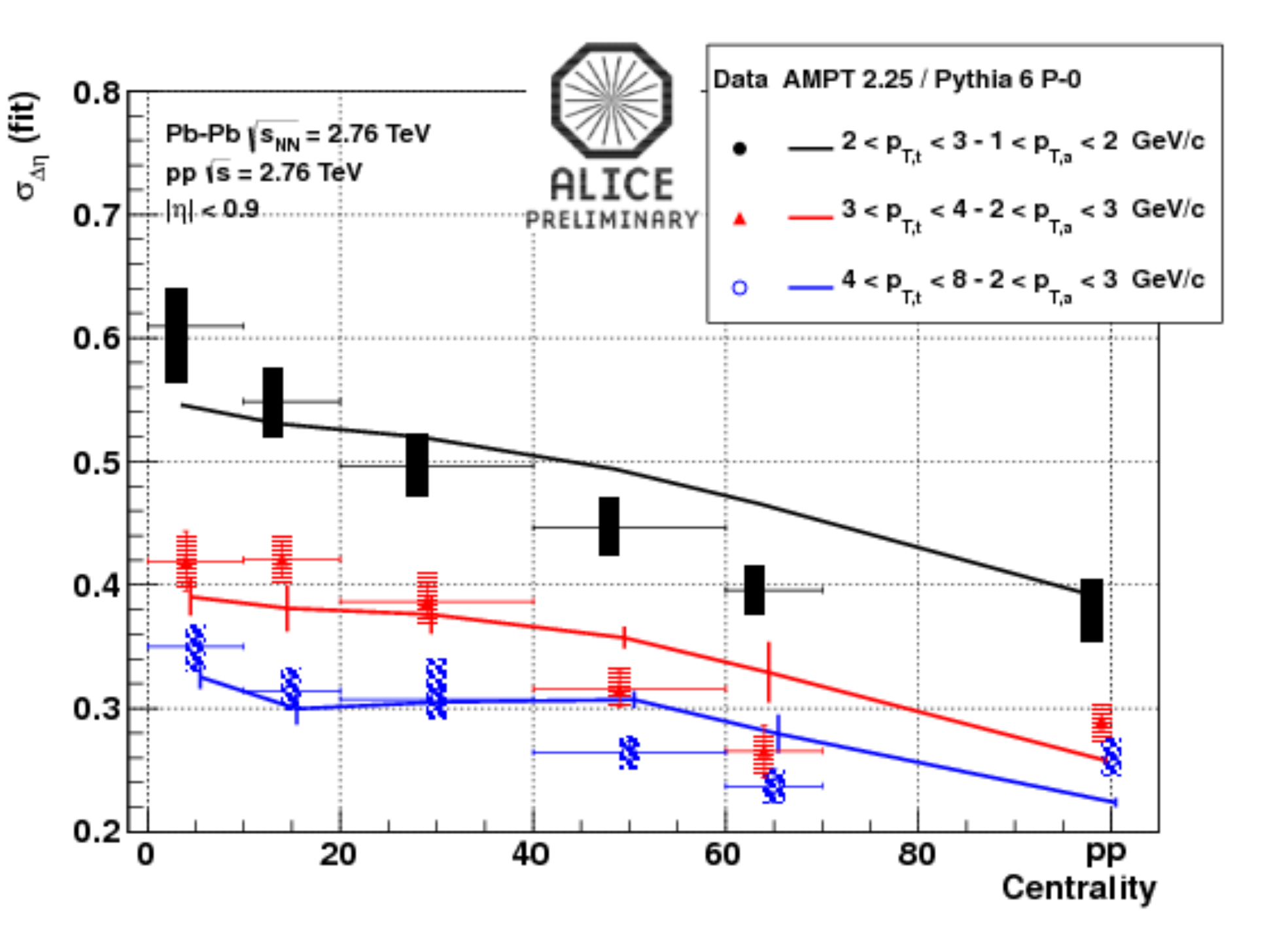}
  \caption{\label{rms} rms in $\Dphi$ (left panel) and $\Deta$ (right panel) calculated from the fit parameters for different $\pt$ bins as function of centrality. Shown are data (points) and AMPT and PYTHIA Monte Carlo simulations (lines).}
\efig

\bfig
  \includegraphics[width=0.48\linewidth]{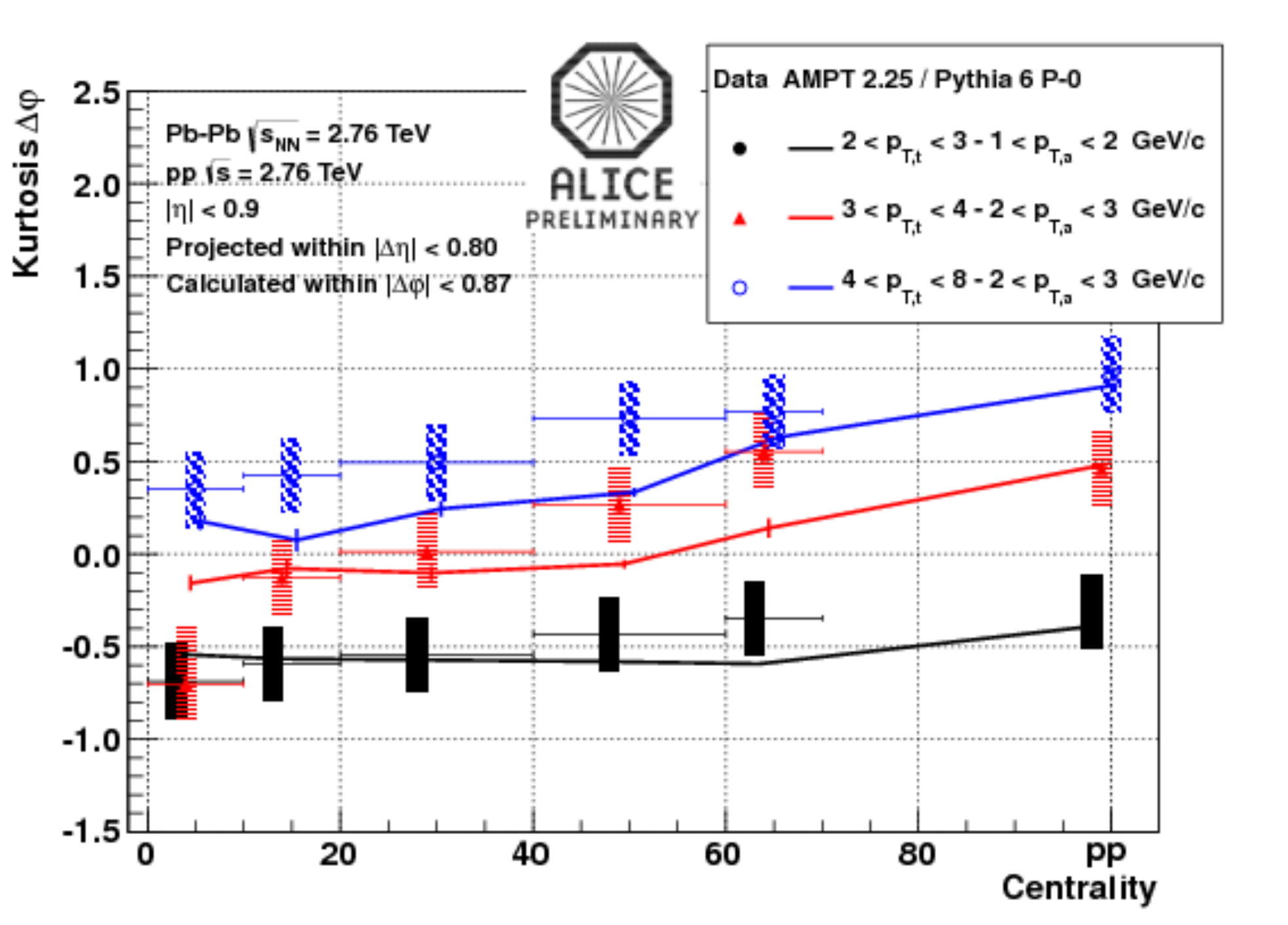}
  \hfill
  \includegraphics[width=0.48\linewidth]{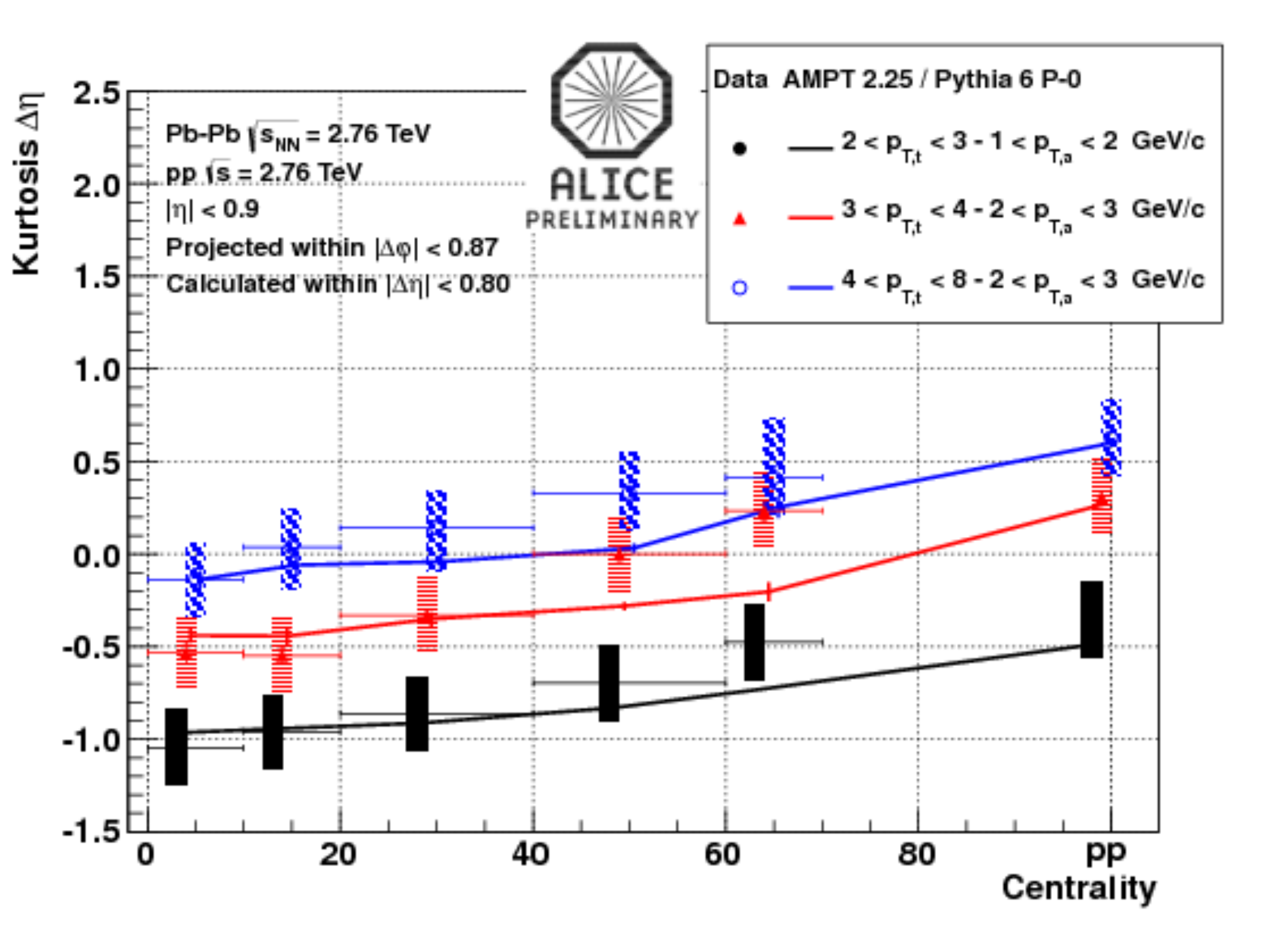}
  \caption{\label{kurtosis} Kurtosis in $\Dphi$ (left panel) and $\Deta$ (right panel) calculated from the fit parameters for different $\pt$ bins as function of centrality. Shown are data (points) and AMPT and PYTHIA Monte Carlo simulations (lines).}
\efig

The lowest $\pt$ bin shows an interesting structure for the most central collisions presented in \figref{lowestbin}: in $\Deta$ direction a departure from a Gaussian is observed. The peak shows a flat top (an indication for a double-humped structure can be seen which is not significant, though). This feature is also reproduced in AMPT. It will be interesting to explore the reasons for such effects in AMPT. More details about this analysis can be found in \cite{andreasproceeding}.

\bfig
  \includegraphics[width=0.66\linewidth,clip=true,trim=190 0 0 20]{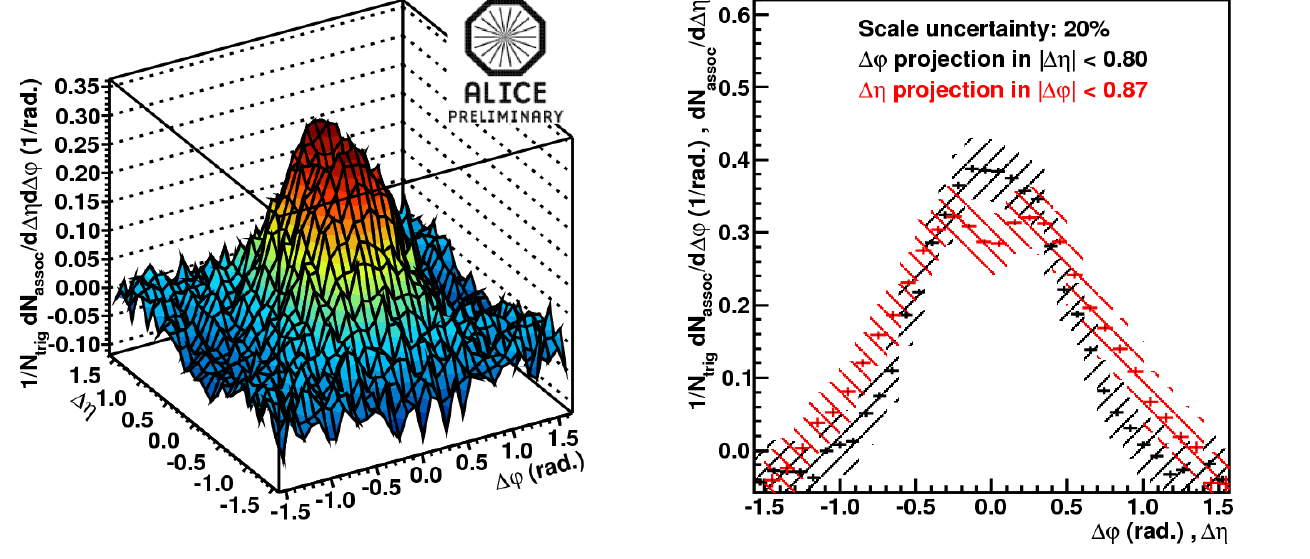}
  \hfill
  \includegraphics[width=0.33\linewidth,clip=true,trim=380 0 0 12]{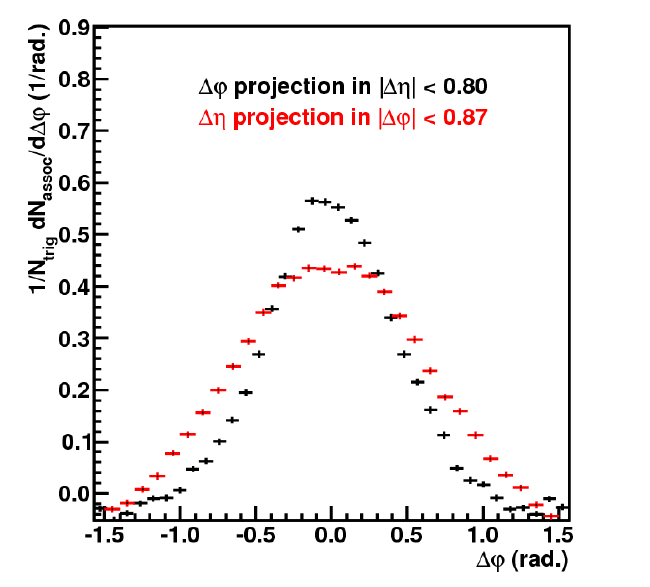}
  \caption{\label{lowestbin} Per-trigger yield for $\unit[2]{GeV/\emph{c}} < \ptt < \unit[3]{GeV/\emph{c}}$, $\unit[1]{GeV/\emph{c}} < \pta < \unit[2]{GeV/\emph{c}}$ and 0-10\% centrality. The left panel shows the data; the middle panel shows projections of the data to $\Dphi$ (black) and $\Deta$ (red) while the right panel shows the same projections from AMPT simulations.}
\efig

\section{p/$\pi$ Ratio in Jet and Bulk}

Baryon over meson ratios differ significantly between central heavy-ion and pp collisions. E.g. the $\Lambda/K^{\rm 0}_{\rm S}$ ratio increases up to about 1.5 in central Pb--Pb collisions at the LHC, while it is about 0.5 in pp collisions \cite{lambdak0ratio}. We study particle ratios in the near-side peak by performing two-particle correlations with identified particles. This allows to separate the bulk of the particles from those associated to a high-$\pt$ trigger particle. 

The particle identification is performed by exploiting specific energy loss as well as time of flight information. The left panel of \figref{pid} shows an example for one $\pt$ bin and the pion hypothesis. The $x$ and $y$ axis show the measured specific energy loss and time of flight, respectively, minus the expected one. For each abundant particle species, pions (around 0, by construction), kaons and protons, a corresponding peak is observed. The pion peak is approximately Gaussian while the others are distorted because the incorrect (pion) mass hypothesis is used also for those particles. These peaks are fitted to extract the particle yields. For the pion peak a Gaussian (plus an exponential tail towards positive $t_{\rm TOF}$) is used while Monte Carlo templates are used for the other ones. 
The center and right panels of \figref{pid} show projections of the data and fit function to the specific energy loss and time of flight axis, respectively. 
This procedure is performed for each particle species, i.e. the corresponding mass hypothesis is used, to extract the yield for that species.
These yields are corrected for tracking and PID efficiency based on Monte Carlo simulations. No correction for feeddown from e.g. $\Lambda$ has been applied.

\bfig
  \includegraphics[width=0.33\linewidth,clip=true,trim=0 12 0 0]{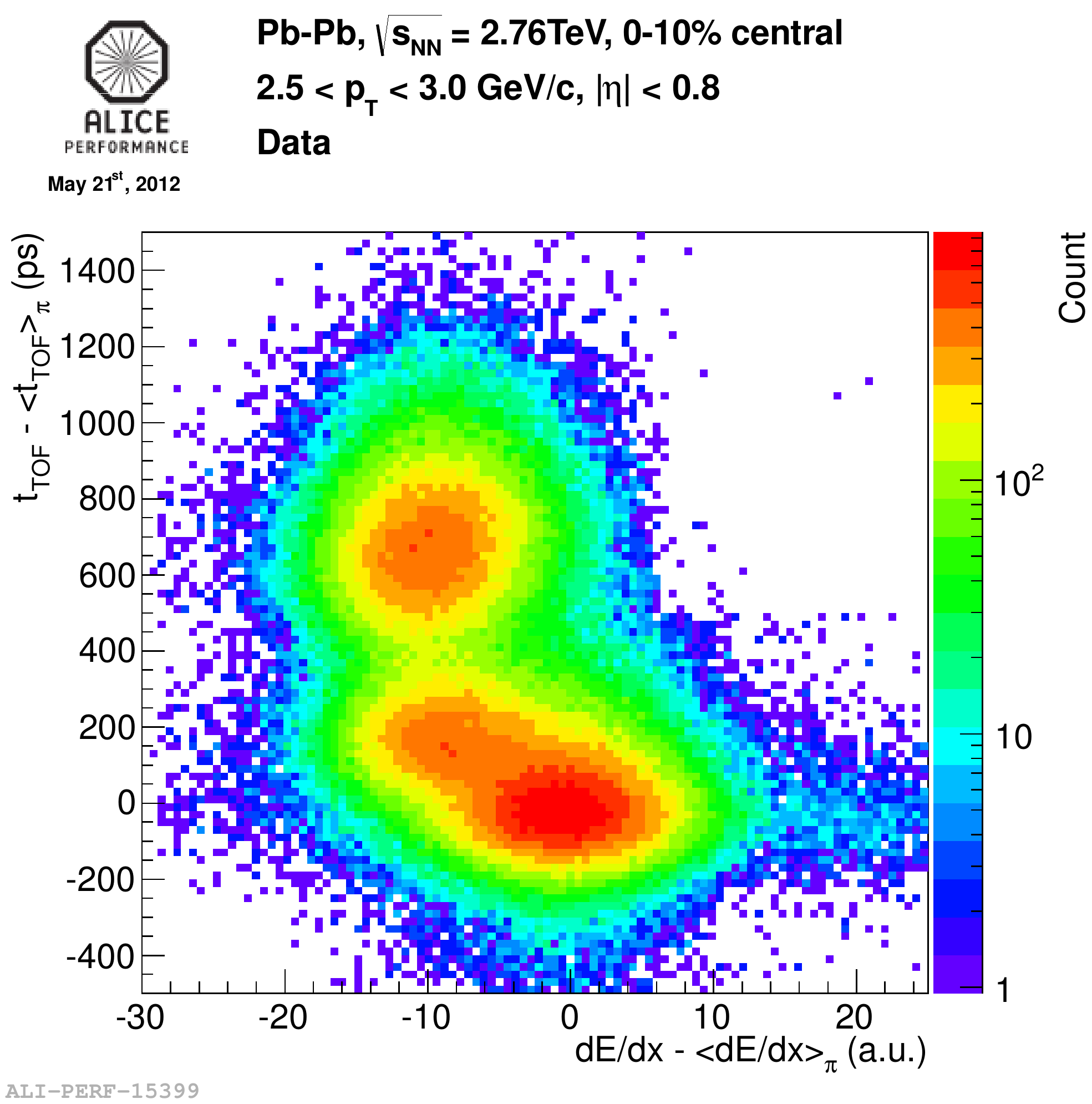}
  \hfill
  \includegraphics[width=0.33\linewidth,clip=true,trim=0 12 0 0]{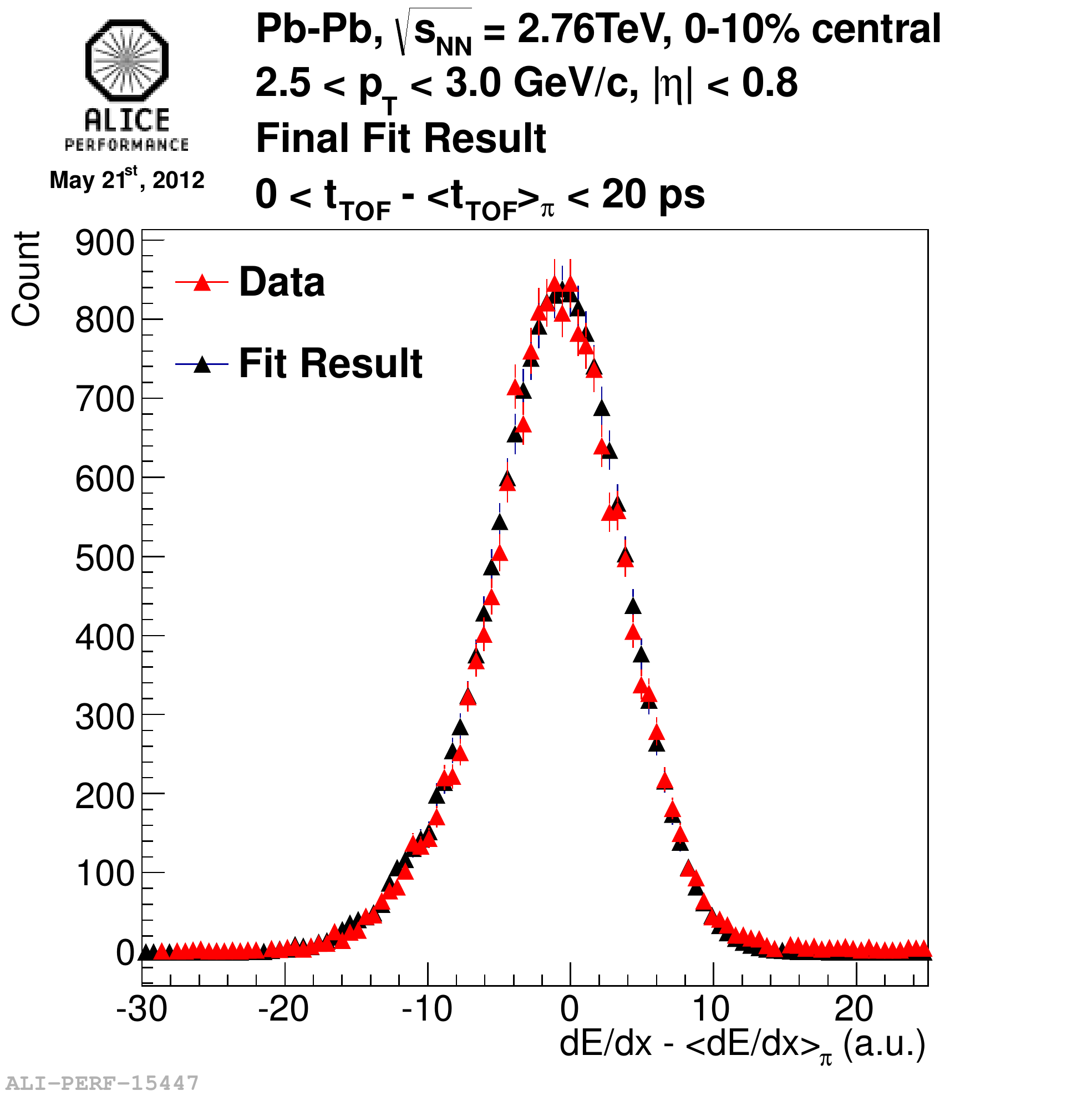}
  \hfill
  \includegraphics[width=0.33\linewidth,clip=true,trim=0 12 0 0]{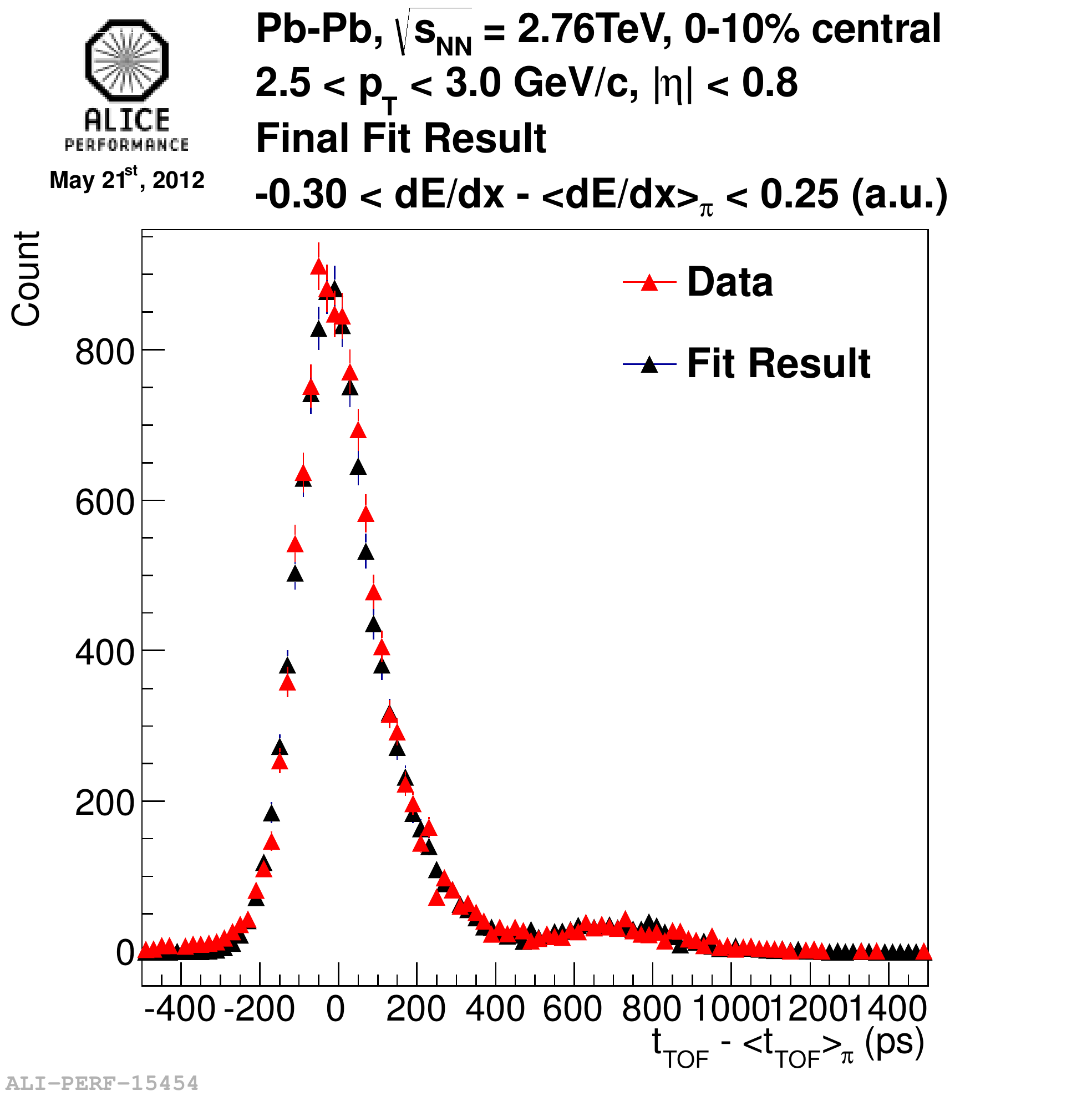}
  \caption{\label{pid} Illustration of the particle identification technique. Left panel: yield of particles as function of measured minus expected time of flight vs measured minus expected specific energy loss; middle panel: projection of the data (black) and the fit function (red) to the specific energy loss axis; right panel: projection of the data (black) and the fit function (red) to the time of flight axis.}
\efig

Similarly to the previous discussed analysis the yields in the long-range correlation region are subtracted from the yields in the near-side peak region. The left panel of \figref{regions} shows the regions defined in the $\Dphi$-$\Deta$ plane: the \emph{bulk I} and \emph{II} regions are used to estimate the background under the \emph{peak region}. The remaining yield in the peak region is the yield correlated to the trigger particle and is called \emph{jet} yield in the following. More details about this analysis can be found in \cite{mishaproceeding}.

\bfig
  \includegraphics[width=0.48\linewidth]{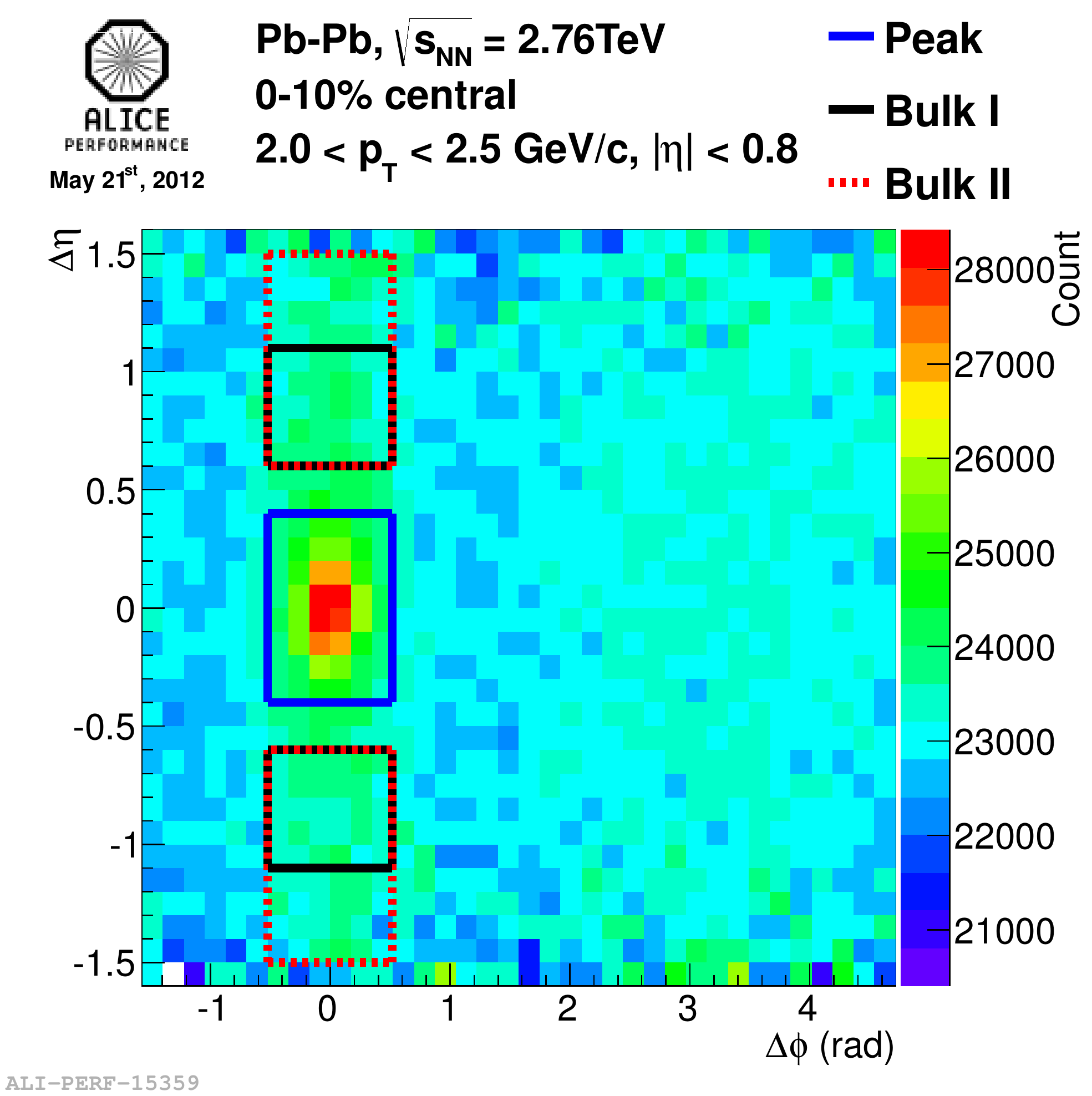}
  \includegraphics[width=0.48\linewidth]{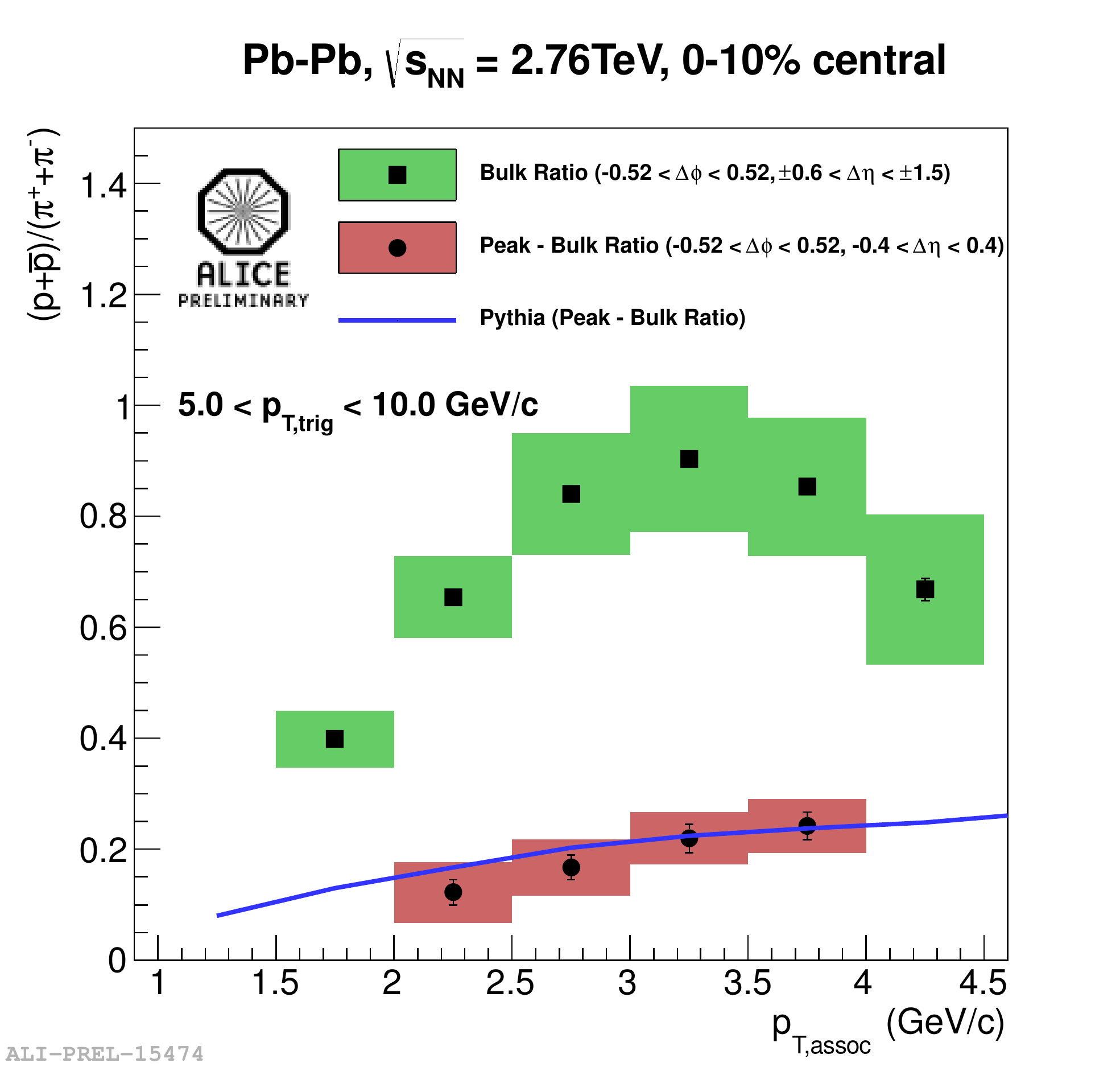}
  \caption{\label{regions} Left panel: regions in the $\Dphi$-$\Deta$ plane where the yields for the peak and bulk are extracted (for details see text). Right panel: $p/\pi$ ratio in the bulk (squares) and peak$-$bulk (circles) regions compared to the PYTHIA expectation (line).}
\efig

The particle yields are used to calculate the $p/\pi$ ratio as a function of $\pta$ for $\unit[5]{GeV/\emph{c}} < \ptt < \unit[10]{GeV/\emph{c}}$, shown in the right panel of \figref{regions}. The $p/\pi$ ratio in the bulk region increases up to about 1 at about \unit[3]{GeV/$c$} and is similar to the inclusive one calculated over all particles (not shown). The ratio of the particles associated to the jet is significantly smaller (maximum at about 0.3) and consistent with the PYTHIA simulation (version 6.4 default tune) which assumes vacuum fragmentation. Hence, this observable shows no evidence for medium-induced modification of the jet fragmentation in central Pb--Pb collisions.

\section{Summary}

Two-particle correlations have been used to quantify the effects of the hot and dense medium on the near-side peak associated to a trigger particle with a transverse momentum in the range \unit[2-10]{GeV/$c$}. The near-side peak shape has been studied revealing that the symmetric peak in peripheral and pp collisions gets asymmetric in central Pb--Pb collisions: the rms in $\Deta$ is significantly larger than in $\Dphi$. Rms and excess kurtosis are well reproduced by AMPT compatible with the interpretation that the interplay of jets with the flowing bulk is the origin of the found feature. The associated $p/\pi$ ratio to a trigger particle has been found to be much smaller than the one in the bulk. The $p/\pi$ ratio is compatible with simulations assuming vacuum fragmentation, i.e. no evidence for medium-induced modification of the jet fragmentation is observed in the studied $\pt$ regime. In summary, two interesting observations have been presented and a continuation as well as combination of these studies seems very promising to gain further insight into the jet quenching mechanism occurring in heavy-ion collisions at the LHC.

%% The Appendices part is started with the command \appendix;
%% appendix sections are then done as normal sections
%% \appendix

%% \section{}
%% \label{}

%% References
%%
%% Following citation commands can be used in the body text:
%% Usage of \cite is as follows:
%%   \cite{key}         ==>>  [#]
%%   \cite[chap. 2]{key} ==>> [#, chap. 2]
%%

%% References with BibTeX database:

%\bibliographystyle{elsarticle-num}
%\bibliography{correlations}

%% Authors are advised to use a BibTeX database file for their reference list.
%% The provided style file elsarticle-num.bst formats references in the required Procedia style

%% For references without a BibTeX database:

% \begin{thebibliography}{00}

\end{document}